\documentclass[preprint,showkeys,prc]{revtex4-1}
\usepackage{epsf}
\usepackage{amsmath,amsthm,amssymb}
\usepackage{color}
\usepackage{dcolumn}
\usepackage{bm}
\usepackage{graphicx,epsfig}

\graphicspath{{fig/}}
\everymath{\displaystyle}

\begin{document}           

\title{General classical and quantum-mechanical description of magnetic resonance:
An application to electric-dipole-moment experiments}

\author{Alexander J. Silenko} 
\affiliation{Research Institute for Nuclear Problems, Belarusian State
University, Minsk 220030, Belarus} \affiliation{
Bogoliubov Laboratory of Theoretical Physics, Joint Institute for Nuclear Research, Dubna 141980, Russia}

\date{\today}

\begin {abstract}
A general theoretical description of a magnetic resonance is presented. This description is necessary for a detailed analysis of spin dynamics in electric-dipole-moment experiments in storage rings. General formulas describing a behavior of all components of the polarization vector at the magnetic resonance are obtained for an arbitrary initial polarization. These formulas are exact on condition that the nonresonance rotating field is neglected. The spin dynamics is also calculated at frequencies far from resonance with allowance for both rotating fields. A general quantum-mechanical analysis of the spin evolution at the magnetic resonance is fulfilled and the full agreement between the classical and quantum-mechanical approaches is shown. Quasimagnetic resonances for particles and nuclei moving in noncontinuous perturbing fields of accelerators and storage rings are considered. Distinguishing features of quasimagnetic resonances in storage ring electric-dipole-moment experiments are investigated in detail. The exact formulas for the effect caused by the electric dipole moment are derived. The difference between the resonance effects conditioned by the rf electric-field flipper and the rf Wien filter is found and is calculated for the first time. The existence of this difference is crucial for the establishment of a consent between analytical derivations and computer simulations and for checking spin tracking programs. Main systematical errors are considered.
\end{abstract}

\maketitle

\section{Introduction}

The magnetic resonance (MR) is a powerful tool of investigation of basic properties of particles and nuclei. The MR is also successfully used for studies of atoms in condensed matters. The theory of the MR is presented in many books (see, e.g., Refs. \cite{Slichter,MLevitt}) and research articles. As a rule, the MR is a spin resonance and its classical and quantum-mechanical descriptions are equivalent. In the present work, we rigorously prove the equivalence of these descriptions in the general case and apply the general theory for an analysis of resonance effects in electric-dipole-moment experiments with polarized beams in storage rings.

The use of the MR in nuclear, particle, atomic, and condensed matter physics consists in a determination of a spin deflection from the initial vertical direction. To find the magnetic moment, one measures dynamics of the vertical polarization. An exhaustive analysis of the spin evolution in storage ring electric-dipole-moment (EDM) experiments needs an advanced description of magnetic and quasimagnetic resonances. In this case, spin interactions of moving particles and nuclei with magnetic and electric fields are defined by the Thomas-Bargmann-Mishel-Telegdi (T-BMT) equation \cite{Thomas-BMT} or by its extension taking into account the EDM \cite{NKFS,GBMT,PhysScr}. An investigation of quasimagnetic resonances caused by the EDM is important for planned experiments with a rf electric-field flipper and a rf Wien filter (see Ref. \cite{rfWienFilter,Mey,rfWienFilterNIMA}). A change of the spin (pseudo)vector is orthogonal to the spin direction. One needs therefore to measure minor (horizontal) polarization components when a resonance is stimulated by a comparatively weak interaction. In particular, this situation takes place for a search for EDMs. In storage ring experiments, it can be convenient to use an initial horizontal beam polarization and to measure an evolution of the vertical spin component. A needed experimental precision is very high. In addition, the resonance fields of the rf electric-field flipper and the rf Wien filter are noncontinuous. For these reasons, general formulas describing spin dynamics at the magnetic and quasimagnetic resonances and their specific application are necessary.

In Sec. \ref{ClaMres}, we give the general description of the MR in the framework of classical spin physics and electrodynamics. The problem of the spin evolution at frequencies far from resonance is solved in Sec. \ref{far}. The general quantum-mechanical description of spin dynamics at the MR is presented in Sec. \ref{Quantum-mechanical}.
Section \ref{NuclPartMotion} is devoted to a discussion of the magnetic and quasimagnetic resonances for moving particles and nuclei. Quasimagnetic resonances for particles and nuclei moving in noncontinuous perturbing fields of accelerators and storage rings are considered in Sec. \ref{discontinuity}. Distinguishing features of quasimagnetic resonances in storage ring electric-dipole-moment experiments are investigated in Sec. \ref{Distinguishing}. Finally, we summarize the obtained results in Sec. \ref{summary}.

The system of units $\hbar=1,~c=1$ is used. We include $\hbar$ and
$c$ into some equations when this inclusion clarifies the problem.

\section{General classical description of nuclear magnetic resonance}\label{ClaMres}

In this section, we consider a usual design of the MR and obtain general equations describing the spin dynamics. While the results obtained are mostly known, the presented study allows us to apply a common approach for a consideration of classical and quantum-mechanical effects.

Let a spinning nucleus be placed into the magnetic field $\bm B_0=B_0\bm e_z$ and the angular frequency of the spin rotation is equal to $\omega_0$. In this case, a rotating or oscillating horizontal magnetic field with a closed angular frequency $\omega\approx\omega_0$ can significantly deflect the nucleus spin from the initial vertical direction. In particular, this effect allows one to measure magnetic moments of nuclei/particles with a high precision.

As a rule, one applies the main vertical magnetic field $\bm{B}_0$ and the oscillating horizontal magnetic field $\bm{\mathcal{B}}\cos{(\omega t+\chi)}$.  In this case, the spin-dependent part of the classical Hamiltonian is given by
\begin{equation}
\begin{array}{c}
H=\bm\omega_0\cdot\bm\zeta+2\bm{\mathfrak{E}}\cdot\bm\zeta\cos{(\omega t+\chi)},~~~ \bm\omega_0=-\frac
{g_N\mu_N}{\hbar}\bm{B}_0,\\ \bm{\mathfrak{E}}=-\frac{g_N\mu_N}{2\hbar}\bm{\mathcal{B}},
\end{array}
\label{propt}\end{equation}
where $\bm\omega_0$ is the angular velocity of the spin precession at the absence of the horizontal magnetic field, $\bm\zeta$ is the spin (pseudo)vector, $g_N$ is the nuclear $g$-factor, and $\mu_N$ is the nuclear magneton. For particles, $g_N\mu_N$ should be replaced with $eg\hbar/(2m)$, where $g=2mc\mu/(es)$.


The direction of the (pseudo)vector $\bm\omega_0$ defines the orientation of the so-called stable spin axis. In the absence of oscillating fields, the spin remains stable if it is initially aligned along this direction. If the initial spin orientation is different, the spin describes a cone around the direction of $\bm\omega_0$. The stable spin axis is a static quantity defined \emph{before} activating the rf.

It is preferable to decompose the oscillating horizontal magnetic field into two magnetic fields rotating in opposite directions. The amplitudes of the rotating magnetic fields are equal to $\bm{\mathcal{B}}/2$. We suppose that $\omega$ is close to $\omega_0$. In this case, one rotating field is resonant and an effect of another rotating field can be neglected. Let us direct $\bm{\mathfrak{E}}$ along the $x$ axis:
\begin{equation}\bm{\mathfrak{E}}=\mathfrak{E}\bm e_x.\label{rvnie} \end{equation}
This direction is not important in the considered case. The turn of the direction of $\bm{\mathfrak{E}}$ by the angle $\varphi$ is equivalent to the change of phase of the rotating resonant field by the same angle.


To calculate the spin dynamics, it is convenient to use the frame rotating about the $z$ axis with the angular velocity $\bm\omega$. We suppose that the direction of the frame rotation coincides with the direction of the spin rotation, $\bm\omega_0$. The horizontal magnetic field rotating in the lab frame becomes constant in the rotating frame. In this frame, the spin rotates about the $z$ axis with the angular frequency $\omega_0-\omega$ and the total angular velocity of the spin rotation is equal to
\begin{equation}
\begin{array}{c}
\bm\Omega=\bm\omega_0-\bm\omega+ \bm{\mathfrak{E}} , ~~~\Omega=\sqrt{(\omega_0-\omega)^2+\mathfrak{E}^2}.
\end{array}
\label{anvelom} \end{equation}
We disregard the rotating field which rotation direction is opposite to that of the spin. The vectors $\bm\omega_0$ and $\bm\omega$ are parallel and $\bm{\mathfrak{E}}$ is the vector with the constant module $\mathfrak{E}$ rotating in the horizontal plane.

It is convenient to use the rotating coordinate system which axis $\bm e'_z$ coincides with the axis $\bm e_z$ of the lab frame. The axis $\bm e'_x$ is supposed to be parallel to the direction of the rotating field. The connection between the coordinate axes of the two frames is given by
\begin{equation}
\begin{array}{c}
\bm e_x=\cos{(\omega t+\chi)}\bm e'_x-\sin{(\omega t+\chi)}\bm e'_y,~~~ \bm e_y=\sin{(\omega t+\chi)}\bm e'_x+\cos{(\omega t+\chi)}\bm e'_y,\\ \bm e_z=\bm e'_z.
\end{array}
\label{contd} \end{equation}

It is also convenient to introduce another rotating coordinate system and to direct its vertical axis, $\bm e''_z$, along the vector $\bm\Omega$. Since this vector lies in the plane $x'z'$,
\begin{equation}
\begin{array}{c}
\bm e''_x=\frac{\omega_0-\omega}{\Omega}\bm e'_x-\frac{\mathfrak{E}}{\Omega}\bm e'_z,~~~ \bm e''_y=\bm e'_y,~~~
\bm e''_z=\frac{\mathfrak{E}}{\Omega}\bm e'_x+\frac{\omega_0-\omega}{\Omega}\bm e'_z.
\end{array}
\label{condp} \end{equation}

The spin rotates about the axis $\bm e_z''$ with the angular frequency $\Omega$. As a result, the component $P_z''$ of the unit polarization vector is constant. Other components are given by
\begin{equation}
\begin{array}{c}
P''_x(t)=P''_x(0)\cos{\Omega t}-P''_y(0)\sin{\Omega t},~~~
P''_y(t)=P''_x(t)\sin{\Omega t}+P''_y(t)\cos{\Omega t}.
\end{array}
\label{pondp} \end{equation}

Equations (\ref{condp}) and (\ref{pondp}) allow us to find dynamics of the polarization vector in the primed coordinate system:
\begin{equation}
\begin{array}{c}
P'_x(t)=\left[1-\frac{(\omega_0-\omega)^2}{\Omega^2}\left(1-\cos{\Omega t}\right)\right]P'_x(0)-\frac{\omega_0-\omega}{\Omega}\sin{\Omega t}P'_y(0)\\+\frac{(\omega_0-\omega)\mathfrak{E}}{\Omega^2}\left(1-\cos{\Omega t}\right)P'_z(0),\\
P'_y(t)=\left[\frac{\omega_0-\omega}{\Omega}P'_x(0)-\frac{\mathfrak{E}}{\Omega}P'_z(0)\right]\sin{\Omega t}+\cos{\Omega t}P'_y(0),\\
P'_z(t)=\frac{(\omega_0-\omega)\mathfrak{E}}{\Omega^2}\left(1-\cos{\Omega t}\right)P'_x(0)+\frac{\mathfrak{E}}{\Omega}\sin{\Omega t}P'_y(0)\\+\left[1-\frac{\mathfrak{E}^2}{\Omega^2}\left(1-\cos{\Omega t}\right)\right]P'_z(0).
\end{array}
\label{pfndp} \end{equation}

We can use now Eq. (\ref{contd}) and express components of the polarization vector in the primed coordinate system in terms of $P_i(t),~P_i(0)~ (i=x,y,z)$.
When the initial spin direction is defined by the spherical angles $\theta$ and $\psi$,
\begin{equation}
P_x(0)=\sin{\theta}\cos{\psi},~~~P_y(0)=\sin{\theta}\sin{\psi},~~~P_z(0)=\cos{\theta}, \label{pfndppfndp} \end{equation}
the final result is given by
\begin{equation}
\begin{array}{c}
P_x(t)=\cos{\Omega t}\sin{\theta}\cos{(\omega t+\psi)}+\frac{\mathfrak{E}^2}{\Omega^2}\left(1-\cos{\Omega t}\right)\sin{\theta}\cos{(\psi-\chi)}\cos{(\omega t+\chi)}\\-\frac{\omega_0-\omega}{\Omega}\sin{\Omega t}\sin{\theta}\sin{(\omega t+\psi)}\\+\frac{\mathfrak{E}}{\Omega}\left[\frac{\omega_0-\omega}{\Omega}\left(1-\cos{\Omega t}\right)\cos{(\omega t+\chi)}+\sin{\Omega t}\sin{(\omega t+\chi)}\right]\cos{\theta},\\
P_y(t)=\frac{\omega_0-\omega}{\Omega}\sin{\Omega t}\sin{\theta}\cos{(\omega t+\psi)}+\cos{\Omega t}\sin{\theta}\sin{(\omega t+\psi)}\\+\frac{\mathfrak{E}^2}{\Omega^2}\left(1-\cos{\Omega t}\right)\sin{\theta}\cos{(\psi-\chi)}\sin{(\omega t+\chi)}\\+\frac{\mathfrak{E}}{\Omega}\left[\frac{\omega_0-\omega}{\Omega}\left(1-\cos{\Omega t}\right)\sin{(\omega t+\chi)}-\sin{\Omega t}\cos{(\omega t+\chi)}\right]\cos{\theta},\\
P_z(t)=\frac{(\omega_0-\omega)\mathfrak{E}}{\Omega^2}\left(1-\cos{\Omega t}\right)\sin{\theta}\cos{(\psi-\chi)}+\frac{\mathfrak{E}}{\Omega}\sin{\Omega t}\sin{\theta}\sin{(\psi-\chi)}\\+\left[1-\frac{\mathfrak{E}^2}{\Omega^2}\left(1-\cos{\Omega t}\right)\right]\cos{\theta}.
\end{array}
\label{findp} \end{equation}

Equation (\ref{findp}) presents the general classical description of spin dynamics at magnetic and quasimagnetic resonances and allows us to conclude that one can use both vertical and horizontal initial polarizations.
When the terms proportional to $\mathfrak{E}^2$ are neglected, Eq. (\ref{findp}) takes the form
\begin{equation}
\begin{array}{c}
P_x(t)=\sin{\theta}\cos{(\omega_0 t+\psi)}\\+\frac{\mathfrak{E}}{\omega_0 -\omega}\biggl\{\cos{(\omega t+\chi)}\left[1-\cos{(\omega_0 -\omega)t}\right]+\sin{(\omega t+\chi)}\sin{(\omega_0 -\omega)t}\biggr\}\cos{\theta},\\
P_y(t)=\sin{\theta}\sin{(\omega_0 t+\psi)}\\+\frac{\mathfrak{E}}{\omega_0 -\omega}\biggl\{\sin{(\omega t+\chi)}\left[1-\cos{(\omega_0 -\omega)t}\right]-\cos{(\omega t+\chi)}\sin{(\omega_0 -\omega)t}\biggr\}\cos{\theta},\\
P_z(t)=\cos{\theta}\\+\frac{\mathfrak{E}}{\omega_0 -\omega}\biggl\{\left[1-\cos{(\omega_0 -\omega)t}\right]\cos{(\psi-\chi)}+\sin{(\omega_0 -\omega)t}\sin{(\psi-\chi)}\biggr\}\sin{\theta}.
\end{array}
\label{polvmop} \end{equation}

\section{Spin evolution at frequencies far from resonance}\label{far}

It is instructive to consider the spin evolution at frequencies far from resonance. This is especially important for the storage ring electric-dipole-moment experiments because some periodical perturbations may imitate the presence of an EDM.

In the considered case, effects of two magnetic fields rotating in opposite directions are comparable. Therefore, it is not appropriate to decompose $\bm{\mathcal{B}}$ in Eq. (\ref{propt}) into the two rotating fields.

It is significant that the results presented in this section are also applicable to a horizontal perturbation caused by a \emph{constant} field. In this case, $\omega=0$.

Let us direct the $x$ axis along the vectors $\bm{\mathcal{B}}$ and $\bm{\mathfrak{E}}$. 
When the perturbation is negligible, the spin rotates with the angular velocity $\bm\omega_0$. Therefore, it is convenient to use the primed coordinate system rotating with this angular velocity:
\begin{equation}
\begin{array}{c}
\bm e'_x=\cos{(\omega_0 t)}\bm e_x+\sin{(\omega_0 t)}\bm e_y,~~~ \bm e'_y=-\sin{(\omega_0 t)}\bm e_x+\cos{(\omega_0 t)}\bm e_y,~~~ \bm e'_z=\bm e_z.
\end{array}
\label{contWvw} \end{equation}

If we denote $\bm{\mathfrak{K}}=2\bm{\mathfrak{E}}\cos{(\omega t+\chi)}=2\mathfrak{E}\cos{(\omega t+\chi)}\bm e_x$, the spin dynamics in the primed frame is defined by
\begin{equation}
\begin{array}{c}
\frac{d\bm\zeta'}{dt}=\bm{\mathfrak{K}}'\times\bm\zeta', ~~~ \mathfrak{K}_x'= 2\mathfrak{E}\cos{(\omega t+\chi)}\cos{(\omega_0 t)},\\
\mathfrak{K}_y'=-2\mathfrak{E}\cos{(\omega t+\chi)}\sin{(\omega_0 t)}.
\end{array}
\label{spinmon} \end{equation}

It is convenient to present the unit spin vector as a sum of two parts, $\bm\zeta(t)=\bm{\mathcal S}(t)+\bm\eta(t)$, where $\bm{\mathcal S}$ rotates with the angular velocity $\bm\omega_0$.
In this case, $\bm{\mathcal S}'$ is constant, $\bm{\mathcal S}'=\bm\zeta'(0)$, and $\bm\eta(0)=0$. In the used approximation,
\begin{equation}
\begin{array}{c}
\frac{d\bm\zeta'}{dt}=\frac{d\bm\eta'}{dt}=\bm{\mathfrak{K}}'\times\bm{\mathcal S}'.
\end{array}
\label{spinmno} \end{equation}

In the general case, the initial spin direction is defined by the spherical angles $\theta$ and $\psi$ [see Eq. (\ref{pfndppfndp})].
An integration on time results in the following evolution of the polarization vector $\bm P=\bm\zeta/s$:
\begin{equation}
\begin{array}{c}
P_x(t)=\sin{\theta}\cos{(\omega_0 t+\psi)}\\+\Biggl(\frac{\mathfrak{E}}{\omega_0 +\omega}\biggl\{\cos{(\omega t+\chi)}\left[1-\cos{(\omega_0 +\omega)t}\right]-\sin{(\omega t+\chi)}\sin{(\omega_0 +\omega)t}\biggr\}\\+\frac{\mathfrak{E}}{\omega_0 -\omega}\biggl\{\cos{(\omega t+\chi)}\left[1-\cos{(\omega_0 -\omega)t}\right]+\sin{(\omega t+\chi)}\sin{(\omega_0 -\omega)t}\biggr\}\Biggr)\cos{\theta},\\
P_y(t)=\sin{\theta}\sin{(\omega_0 t+\psi)}\\+\Biggl(-\frac{\mathfrak{E}}{\omega_0 +\omega}\biggl\{\sin{(\omega t+\chi)}\left[1-\cos{(\omega_0 +\omega)t}\right]+\cos{(\omega t+\chi)}\sin{(\omega_0 +\omega)t}\biggr\}\\+\frac{\mathfrak{E}}{\omega_0 -\omega}\biggl\{\sin{(\omega t+\chi)}\left[1-\cos{(\omega_0 -\omega)t}\right]-\cos{(\omega t+\chi)}\sin{(\omega_0 -\omega)t}\biggr\}\Biggr)\cos{\theta},\\
P_z(t)=\cos{\theta}\\+\Biggl(\frac{\mathfrak{E}}{\omega_0 +\omega}\biggl\{\left[1-\cos{(\omega_0 +\omega)t}\right]\cos{(\psi+\chi)}+\sin{(\omega_0 +\omega)t}\sin{(\psi+\chi)}\biggr\}\\ +\frac{\mathfrak{E}}{\omega_0 -\omega}\biggl\{\left[1-\cos{(\omega_0 -\omega)t}\right]\cos{(\psi-\chi)}+\sin{(\omega_0 -\omega)t}\sin{(\psi-\chi)}\biggr\}\Biggr)\sin{\theta}.
\end{array}
\label{polvmon} \end{equation}
Evidently, this equation is compatible with Eq. (\ref{polvmop}). Equation (\ref{polvmon}) can be reduced to the form
\begin{equation}
\begin{array}{c}
P_x(t)=\sin{\theta}\cos{(\omega_0 t+\psi)}+\Biggl\{\frac{\mathfrak{E}}{\omega_0 +\omega}\biggl[\cos{(\omega t +\chi)}-\cos{(\omega_0t -\chi)}\biggr]\\+\frac{\mathfrak{E}}{\omega_0 -\omega}\biggl[\cos{(\omega t+\chi)}-\cos{(\omega_0t+\chi)}\biggr]\Biggr\}\cos{\theta},\\
P_y(t)=\sin{\theta}\sin{(\omega_0 t+\psi)}+\Biggl\{-\frac{\mathfrak{E}}{\omega_0 +\omega}\biggl[\sin{(\omega t+\chi)}+\sin{(\omega_0t-\chi)}\biggr]\\
+\frac{\mathfrak{E}}{\omega_0 -\omega}\biggl[\sin{(\omega t+\chi)}-\sin{(\omega_0t+\chi)}\biggr]\Biggr\}\cos{\theta},\\
P_z(t)=\cos{\theta}+\Biggl(\frac{\mathfrak{E}}{\omega_0 +\omega}\biggl\{\cos{(\psi+\chi)}-\cos{[(\omega_0 +\omega)t+\psi+\chi]}\biggr\}\\+\frac{\mathfrak{E}}{\omega_0 -\omega}\biggl\{\cos{(\psi-\chi)}-\cos{[(\omega_0 -\omega)t+\psi-\chi]}\biggr\}\Biggr)\sin{\theta}.
\end{array}\label{polvmol} \end{equation}
The result obtained can be useful when a background is known. In this case, measurements at frequencies far from resonance can be applied in a statistical analysis. Otherwise, Eqs. (\ref{polvmon}) and (\ref{polvmol}) make it possible to take into account some periodical perturbations imitating the presence of the EDM. The beam rotates in a storage ring with the cyclotron frequency $\omega_c$. In this case, any local radial or longitudinal magnetic field becomes a perturbation oscillating with the frequency $\omega_c$ or with a multiple frequency in the presence of an appropriate symmetry. Such perturbations are clearly manifested by spin tracking \cite{tracking}.

An analysis of Eqs. (\ref{pfndp}), (\ref{findp}), and (\ref{polvmol}) displays an impotrant property of magnetic and quasimagnetic resonances.
When the perturbing field is rather weak ($|\mathfrak{E}|\ll\Omega$) and it needs to be determined, it is preferable to measure \emph{minor} spin components.
Specifically, the use of the initial vertical polarization requires a measurement of a horizontal polarization. It is of the order of $|\mathfrak{E}|/\Omega$, while a change of the vertical spin component is of the order of ${\mathfrak{E}}^2/\Omega^2$. Both of the horizontal spin components oscillate but only the component collinear to $\bm{\mathfrak{E}}$ is nonzero on average. In practice, a detection of the oscillatory spin motion is easier than that of the
small constant part of the horizontal spin component collinear to $\bm{\mathfrak{E}}$.

When the initial polarization is horizontal, it is preferable to measure the vertical polarization which is of the order of $|\mathfrak{E}|/\Omega$. In the equations for the horizontal spin components, terms proportional to $\mathfrak{E}/\Omega$ vanish when the initial vertical polarization is equal to zero.
Thus, monitoring of these components is not useful.
This approach has been applied for a search for a muon EDM \cite{MuEDM08} in framework of the muon \emph{g}$-$2 experiment at Brookhaven National Laboratory (see Ref. \cite{PRDfinal}).  The search has been carried out for the initial longitudinal polarization of muons and the \emph{constant} radial perturbing field caused by the EDM. The vertical spin component has been detected. This experiment has allowed one to obtain an upper bound on the muon EDM \cite{MuEDM08}.

\section{Quantum-mechanical description of 
magnetic resonance}\label{Quantum-mechanical}

The detailed classical description of the MR given in the previous sections exhaustively defines
the spin motion caused by interactions linear in the spin. This description is well-substantiated because it is based on manifestly covariant initial equations. However, spin-dependent parts of Hamiltonians contain also terms quadratic in the spin for deuteron and other nuclei with the spin $s\ge1$. A resonance experiment for the deuteron ($s=1$) is a part of the EDM program \cite{Lehrach:2012eg,RathmannWienFilter}. A presence of the terms quadratic in the spin leads to systematical effects mimicking the EDM under the MR \cite{Bar3,Bar4,PRC,PRC2008,PRC2009,dEDMtensor}. While the classical description of these effects is possible \cite{Bar3,Bar4}, a more
general theory 
which has been developed in Refs. \cite{PRC,PRC2008,PRC2009} is based on relativistic quantum-mechanical Hamiltonians in the Foldy-Wouthuysen representation (see Ref. \cite{FW} and references therein). In the present study, we do not consider the effects nonlinear in the spin. Nevertheless, a need for future investigations stipulates for 
an advanced quantum-mechanical description of the standard MR conditioned by spin interactions \emph{linear} in the spin. To solve this problem, we may use the Pauli spin matrices even for nuclei with the spin $s\ge1$. This possibility is based on 
universal commutation relations for spin components which are satisfied for any spins. The identity of the spin motion of particles with spins 1/2 and 1 near a resonance has been demonstrated in Ref. \cite{CzJP}.

It is convenient to use the matrix Hamiltonian method for a
quantum-mechanical description of the MR. When spin-tensor
interactions are not taken into account, the spin rotation of
nuclei/particles with spin 1/2 and with higher spins is very
similar. Therefore, spin rotation of nuclei/particles with spin
$s\ge1$ can also be described with the Dirac matrices acting on
the two-component spin wave function. We consider the same field
configuration as in Sec. \ref{ClaMres}.

Any Foldy-Wouthuysen Hamiltonian \cite{FW} of a relativistic spin-1/2 particle can be presented in the form
\begin{equation}
{\cal H}_{FW}={\cal H}_0+\frac12\bm\Sigma\cdot\bm\Omega_1+\frac12\bm\Pi\cdot\bm\Omega_2,
\label{condpFW} \end{equation} where ${\cal H}_0$ is the sum of spin-independent terms. Lower spinors of corresponding Foldy-Wouthuysen wave functions are zeroth.
The classical limit of the sum $\bm\Omega=\bm\Omega_1+\bm\Omega_2$ defines the angular velocity of the spin precession. As a rule, its division into two parts mirrors contributions of spin interactions with the electric and magnetic fields.

Averaging the spin-dependent terms with the four-component spin wave functions $\zeta^+=\left(\begin{array}{c} 1 \\ 0 \\
0 \\ 0\end{array}\right)$ and
$\zeta^-=\left(\begin{array}{c} 0 \\ 1
 \\0 \\ 0\end{array}\right)$ corresponding to the spin-up and spin-down states, respectively, results in
\begin{equation}
<i|\bm\Sigma\cdot\bm\Omega_1+\bm\Pi\cdot\bm\Omega_2|j>=(\bm\sigma\cdot\bm\Omega)_{ij}, ~~~ \bm\Omega=\bm\Omega_1+\bm\Omega_2,
\label{coHam} \end{equation}
where $\bm\Omega$ is the operator of the spin precession which classical limit is defined by Eq. (\ref{anvelom}) and $\bm\sigma$ is the Pauli matrix.
Certainly, averaging is fulfilled with both the spin and coordinate wave functions.

As a result, the matrix Hamiltonian takes the form
\begin{equation}
H=E_0+\frac12\bm\sigma\cdot\bm\Omega.
\label{coHao} \end{equation} It acts on the two-component wave function $\Psi(t)=\left(\begin{array}{c}
\Psi_+(t) \\ \Psi_-(t)
\end{array}\right)$.

The components of the polarization vector are defined by
\begin{eqnarray}
P_i =\frac{<S_i>}{s}, ~~~ i=x,y,z, \label{eq1P}\end{eqnarray}
where $S_i$ are corresponding spin matrices and $s$ is the spin
quantum number. Averages of the spin operators $\sigma_i=2S_i$ are expressed by their convolutions with the wave function
$\Psi(t)$.

The corresponding relations for the polarization vector have the form
\begin{equation}
\begin{array}{c}
P_x=C_+C_-^\ast+C_{-}C_+^\ast,~~~
P_y=i(C_+C_-^\ast-C_-C_+^\ast),~~~
P_z=C_+C_+^\ast-C_{-}C_{-}^\ast.
\end{array}
\label{eqpvuhalf}
\end{equation}

The quantity $\bm{\mathfrak{E}}$ is defined by Eq. (\ref{rvnie}).
Since $$\bm{\mathcal{B}}\cos{(\omega t+\chi)}=(\bm{\mathcal{B}}/2)\{\exp{[i(\omega t+\chi)]}+\exp{[-i(\omega t+\chi)]}\},$$ the Hamiltonian (\ref{coHao}) reads
\begin{equation}
H=E_0+\frac12\left(\begin{array}{cc}\omega_0 &\mathfrak{G}\\ \mathfrak{G}&-\omega_0\end{array}\right),~~~\mathfrak{G}=\mathfrak{E}\{\exp{[i(\omega t+\chi)]}+\exp{[-i(\omega t+\chi)]}\}.
\label{clnto} \end{equation}
It is convenient to make the following transformation of the wave function:
\begin{equation}
\Psi(t)=\left(\begin{array}{cc}\exp{[-i(\omega t+\chi)/2]} & 0 \\ 0 &\exp{[i(\omega t+\chi)/2]}\end{array}\right)\exp{(-iE_0 t)}C(t),
\label{wvefn} \end{equation}
It can be easily shown with the use of Eq. (\ref{eqpvuhalf}) that this transformation is equivalent to the transition to the frame rotating with the angular frequency $\omega$. Since
$$\begin{array}{c}
\Psi_+\Psi_-^\ast+\Psi_{-}\Psi_+^\ast=\cos{(\omega t+\chi)}(C_+C_-^\ast+C_{-}C_+^\ast)-\sin{(\omega t+\chi)}i(C_+C_-^\ast-C_-C_+^\ast),\\
i(\Psi_+\Psi_-^\ast-\Psi_{-}\Psi_+^\ast)=\sin{(\omega t+\chi)}(C_+C_-^\ast+C_{-}C_+^\ast)+\cos{(\omega t+\chi)}i(C_+C_-^\ast-C_-C_+^\ast),\\
\Psi_+\Psi_+^\ast-\Psi_{-}\Psi_{-}^\ast=C_+C_+^\ast-C_{-}C_{-}^\ast,\end{array}$$ the connection between components of the polarization vector $\bm P$ calculated with the wave functions $\Psi(t)$ and $C(t)$, respectively, has the form
\begin{equation}
\begin{array}{c}
P_x=\cos{(\omega t+\chi)}P'_x-\sin{(\omega t+\chi)}P'_y,~~~ P_y=\sin{(\omega t+\chi)}P'_x+\cos{(\omega t+\chi)}P'_y,\\ P_z=P'_z.
\end{array}
\label{contdqm} \end{equation} The primed components $P_i'~(i=x,y,z)$ correspond to the wave function $C(t)$.

Evidently, Eqs. (\ref{contd}) and (\ref{contdqm}) fully agree.

The transformation (\ref{wvefn}) brings the equation for the matrix Hamiltonian to the form
\begin{equation}
i\frac{dC(t)}{dt}=\frac12\left(\begin{array}{cc}\omega_0-\omega & \mathfrak{G}\exp{[i(\omega t+\chi)]}  \\ \mathfrak{G}\exp{[-i(\omega t+\chi)]} & -\omega_0+\omega \end{array}\right)C(t).
\label{wvewf} \end{equation}

The terms in Eq. (\ref{wvewf}) oscillating with the angular frequency $2\omega$ can be neglected and this equation takes the form
\begin{equation}
i\frac{dC(t)}{dt}=\frac12\left(\begin{array}{cc}\omega_0-\omega & \mathfrak{E}  \\ \mathfrak{E} & -\omega_0+\omega \end{array}\right)C(t).
\label{wveff} \end{equation}

The solution of Eq. (\ref{wveff}) is given by
\begin{equation}\begin{array}{c}
C_+(t)=\left(\cos{\frac{\Omega t}{2}}-i\frac{\omega_0-\omega}{\Omega}\sin{\frac{\Omega t}{2}}\right)C_+(0)-i\frac{\mathfrak{E}}{\Omega}
\sin{\frac{\Omega t}{2}}C_-(0),\\
C_-(t)=-i\frac{\mathfrak{E}}{\Omega}\sin{\frac{\Omega t}{2}}C_+(0)+\left(\cos{\frac{\Omega t}{2}}+i\frac{\omega_0-\omega}{\Omega}
\sin{\frac{\Omega t}{2}}\right)C_-(0),
\end{array}
\label{wvoln} \end{equation} where $\Omega$ is defined by Eq. (\ref{anvelom}).

If we use Eq. (\ref{wvoln}) for a derivation of $P'_i(t)$ in terms of $P'_i(0)~(i=x,y,z)$, we come to Eq. (\ref{pfndp}). This fact clearly demonstrates the full agreement of results obtained by the classical and quantum-mechanical approaches. Similarly to the precedent section, we can use Eqs. (\ref{pfndp}), (\ref{pfndppfndp}), and (\ref{contdqm}) for a derivation of the general equation (\ref{findp}). Thus, this equation is valid not only in classical spin physics but also in quantum mechanics.

The quantum-mechanical description of the spin evolution at the MR is often used in textbooks (see Ref. \cite{MLevitt}) and research articles. In the present study, the \emph{general} case has been considered and the full agreement between the classical and quantum-mechanical approaches has been demonstrated. This agreement seems to be very natural. However, its proof is not redundant because we should take into account the existence of difference between classical and quantum-mechanical descriptions of some spin effects (see Ref. \cite{Nicolaevici}). In future investigations of resonance phenomena for nuclei with the spin $s\ge1$, taking into account spin interactions quadratic in spin will be necessary. Such interactions are caused by the tensor electric and magnetic polarizabilities and the electric quadrupole moment. In this case, a transition to the spin-1 matrices can be necessary.

\section{Magnetic and quasimagnetic resonances for moving particles and nuclei}\label{NuclPartMotion}

Magnetic and quasimagnetic resonances for moving particles and nuclei have some distinguishing features. The main
difference from the MR for nuclei at rest is the use of the T-BMT equation \cite{Thomas-BMT} or its extension taking into account the EDM \cite{NKFS,GBMT,PhysScr} for a description of spin coupling with external fields. The general equation extended on the EDM defines the angular velocity of spin precession in external electric and magnetic fields in the Cartesian coordinates and has the form \cite{NKFS,GBMT,PhysScr}
\begin{equation}\begin{array}{c}
\frac{d\bm{\zeta}}{dt} =(\bm{\Omega}_{T-BMT}+\bm{\Omega}_{EDM})\times\bm{\zeta},\\ \bm{\Omega}_{T-BMT}=\frac{e}{m}\left[\left(G+\frac{1}{\gamma+1}\right){\bm \beta}\times{\bm E}-
\left(G+\frac{1}{\gamma}\right){\bf B}+\frac{G\gamma}{\gamma+1}({\bm\beta}\cdot{\bm B}){\bm\beta}\right],\\
\bm{\Omega}_{EDM}=-\frac{e\eta}{2m}\left[\bm
E-\frac{\gamma}{\gamma+1}({\bm\beta}\cdot\bm E){\bm\beta}+{\bm\beta} \times\bm
B\right],~~~\bm\beta=\frac{\bf v}{c},
\end{array} \label{eq36cll} \end{equation}
where $G=(g-2)/2,~\eta=2mcd/(es)$, and $d$ is the EDM.

Equation (\ref{eq36cll}) is useful when the fields have definite directions relative to the Cartesian coordinates. As a rule, it is not the case for particles and nuclei in accelerators and storage rings. Their motion is cyclic and the fields are usually orthogonal to the beam trajectory. Therefore, it is natural to
define the fields and the spin motion relative to the radial and longitudinal coordinates. The use of the cylindrical coordinate system \cite{RPJSTAB} decreases the angular frequency of the spin rotation about the vertical axis by the cyclotron frequency $\omega_c$. It is important that the angular frequencies of the spin rotation about the two horizontal axes remain the same. The resulting angular velocity of the spin rotation in the cylindrical coordinate system is equal to \cite{RPJSTAB}
\begin{eqnarray}
\bm\Omega^{(cyl)}=-\frac{e}{m}\left\{G\bm B-
\frac{G\gamma}{\gamma+1}\bm\beta(\bm\beta\cdot\bm B)\right.\nonumber\\
+\left(\frac{1}{\gamma^2-1}-G\right)\left(\bm\beta\times\bm
E\right)+\frac{1}{\gamma}\left[\bm B_\|
-\frac{1}{\beta^2}\left(\bm\beta\times\bm
E\right)_\|\right]\nonumber\\ \left.+ \frac{\eta}{2}\left(\bm
E-\frac{\gamma}{\gamma+1}\bm\beta(\bm\beta\cdot\bm
E)+\bm\beta\!\times\!\bm B\right)\!\right\}. 
\label{eq7}\end{eqnarray}
The sign $\|$ means a horizontal projection for any vector.

As a rule, the vertical and horizontal components of the magnetic and quasimagnetic fields, $\bm B$ and $\bm\beta\times\bm E$, enter into the expressions for $\bm\omega_0$ and $\bm{\mathfrak{E}}$ with different factors. In particular, this situation takes place for a beam in a purely magnetic storage ring:
\begin{equation}
\begin{array}{c}
\bm\omega_0=-\frac
{eG}{m}\bm{B}_0,~~~ \bm{\mathfrak{E}}=-\frac
{e}{2m}\left(\frac{1}{\gamma}
+G\right)\bm{\mathcal{B}}.
\end{array}
\label{proptob}\end{equation} Here $\bm{\mathcal{B}}$ is the amplitude of the perturbing oscillatory magnetic field. In the purely magnetic storage ring, the \emph{average} radial magnetic field is equal to zero. In a storage ring with electric focusing, it should be counterbalanced by a focusing vertical electric field. In this case, the horizontal components of the average Lorentz force vanish:
$$<\bm F_{L\|}>=e\left[<\bm E_\|>+<(\bm\beta\times\bm B)_\|>\right]=0.$$

To describe spin dynamics in accelerators and storage rings, one often uses the Frenet-Serret (FS) coordinate system.
The axes of the FS coordinate system depend on the particle trajectory. Three axes of this coordinate system are directed parallel to the velocity
and momentum, parallel to the acceleration
vector, and along the binormal orthogonal to these two axes.
Relative to the Cartesian
coordinate system, the FS one rotates about all three axes, not only around the vertical
axis as the cylindrical coordinate system.

To find the angular velocity of the spin motion in the FS
coordinate system, it is necessary to subtract an angular velocity of rotation of the vector $\bm N=\bm p/p={\bf v}/v$ from $\bm\Omega_{T-BMT}+\bm{\Omega}_{EDM}$. The angular velocity of the spin motion is equal to \cite{GBMT}
\begin{eqnarray}
\bm\Omega^{(FS)}=-\frac{e}{m}\left[G\bm B-\frac{G\gamma}{\gamma+1}\bm\beta(\bm\beta\cdot\bm B)+\left(\frac{1}{\gamma^2-1}-G\right)\left(\bm\beta\times\bm
E\right)\right.\nonumber\\
+\left.\frac{\eta}{2}\left({\bm E}-\frac{\gamma}{\gamma+1}\bm\beta(\bm\beta\cdot\bm E)+{\bm\beta}\times {\bm B}\right)\right].
\label{Nelsonh}\end{eqnarray}

A comparison of the cylindrical and FS coordinate systems has been made in Ref. \cite{JINRLettCylr}.
Of course, Eq. (\ref{Nelsonh}) is more compact. However, this
compactness is achieved owing to the fact that the directions of the FS coordinate axes change with time, while the directions of the axes of the cylindrical coordinate system are fixed. Therefore, Eq. (\ref{Nelsonh}) can create the illusion that the effect of vertical and radial magnetic fields on the spin is the same. However, when $\bm E=0$ and the EDM is neglected, the ratios $\Omega_\rho^{(cyl)}/B_\rho$ and $\Omega_z^{(cyl)}/B_z$ differ by the factor $(G\gamma+1)/(G\gamma)$.
For leptons (the electron and the muon) this ratio can be
rather large. The reason is that, to determine an \emph{observable} effect, the motion of the axes
of the FS coordinate system should be added to the spin motion in
this coordinate system. Once this circumstance is taken
into account, the cylindrical and FS coordinate systems give an equivalent description of the spin motion \cite{JINRLettCylr}.

In storage ring EDM experiments, the perturbing field should act on the EDM. For this purpose, one can use a resonance radial electric field (a  rf electric-field flipper). One can also add a vertical magnetic field oscillating with the same resonance frequency. When the Lorentz force created by this device is equal to zero, one obtains a rf Wien filter \cite{rfWienFilter}. The very weak perturbing field is caused by the interaction of the EDM with the main magnetic or electric field. This perturbing field is constant and rotates the spin about the radial axis. This case has been discussed at the end of Sec. \ref{far}.

We can mention that a constant perturbation rotating the spin about the radial axis is also conditioned by the Earth's gravity \cite{PRD,PRD2007,OFSgravity}. This perturbation is very weak.

A detailed consideration of evolution of all spin components is necessary because an interaction stimulating a resonance can be very weak. Evidently,  $d\bm\zeta/(dt)\bot\bm\zeta$. When the initial beam polarization is vertical, one needs to measure horizontal spin components. When the initial beam polarization is horizontal, it is convenient to monitor the vertical spin component (see Sec. \ref{far}). These two possibilities may be realized in storage ring experiments on a search for EDMs \cite{OMS,PRC2009,Lehrach:2012eg,JETPLet}. In these cases, the action of oscillating fields on the EDM stimulates a resonance while their action on the magnetic moment does not bring any resonance effects.
This takes place because the quantities $\bm{\Omega}_{T-BMT}$ and $\bm{\Omega}_{EDM}$ in Eq. (\ref{eq36cll}) are usually orthogonal. Corresponding quantities in Eqs. (\ref{eq7}) and (\ref{Nelsonh}) possess the same property. In any case, spin resonances originated from the EDM are quasimagnetic and are not magnetic.
We can also mention that the case when the $x$ and $z$ components of the angular velocity of spin precession are constant [see Eq. (\ref{pfndp})] corresponds to the conditions of the EDM experiment based on the frozen spin method \cite{FJM} (cf. Eq. (22) in Ref. \cite{PRC2009}).

It has been proven in Ref. \cite{JETPLet} that the use of the initial vertical polarization cancels some systematical errors. The use of the initial horizontal polarization does not lead to such a cancellation. However, the initial vertical polarization can meet other problems \cite{JETPLet}.

We can conclude that specific conditions of the magnetic and quasimagnetic resonances for particles and nuclei moving in accelerators and storage rings influence only parameters $\omega_0$ and $\mathfrak{E}$ but do not change the general equations (\ref{findp}) and (\ref{polvmol}) defining the spin dynamics. A calculation of small corrections appearing in exact solutions needs a modification of initial equations. This problem will be considered in the next section.

It can be added that an extremely high precision of storage ring EDM experiments needs taking into account tensor electric and magnetic polarizabilities for nuclei with spin $s\ge1$ (e.g., deuteron) \cite{Bar3}. The tensor magnetic polarizability,
$\beta_T$, produces the spin rotation with two
frequencies instead of one, beating with a frequency proportional to $\beta_T$, and causes
transitions between vector and tensor polarizations
\cite{Bar3,Bar4,PRC2008}. A beam with an initial tensor polarization
acquires a final vector polarization \cite{PRC2009,PRC2008,PRC}. Resonance effects caused by the tensor polarizabilities have been calculated in Refs. \cite{Bar3,Bar4,PRC}. A comparison of spin dynamics conditioned by the tensor polarizabilities and the EDM has been carried out in Refs. \cite{PRC2009,PRC,dEDMtensor}. The corresponding spin motion without taking into account spin-tensor effects is presented by formulas of Sec. \ref{far} provided that $\omega=0$.

\section{Quasimagnetic resonance in a noncontinuous perturbing field}\label{discontinuity}

The next problem which should be taken into consideration in connection with the storage ring EDM experiments is a discontinuity of perturbing fields.
In the planned EDM experiments with protons, deuterons, and $^3$He ions at COSY \cite{Lehrach:2012eg,RathmannWienFilter}, one will use resonance stimulations with a rf electric-field flipper and a rf Wien filter.
The both devices create oscillatory fields. The frequencies of the perturbing fields are synchronized with that of the spin frequency. The both devices provide for standard conditions of the MR. Semertzidis \cite{SemertzidisRFE} and Nikolaev \cite{NikolaevRFE} have compared the actions of the rf electric-field flipper and the rf Wien filter on the spin.
Orlov has shown \cite{OrlovPartiallyFrozenSpin} that a part of the longitudinal spin component is frozen (constant in time) in the oscillatory radial electric field. A more advanced theoretical analysis has been fulfilled in Ref. \cite{rfWienFilter}. The theoretical calculations agree with spin tracking \cite{SemertzidisRFE,rfWienFilter}.

The rf Wien filter unlike the rf electric-field flipper does not affect the motion of particles and nuclei. This is a great advantage of the former device. The latter device can be used only if it does not destroy the beam stability.

JEDI collaboration plans to perform main experiments with the rf Wien filter \cite{RathmannWienFilter,Mey,rfWienFilterNIMA}.
The static version of this filter is frequently used to turn the spin without an effect on beam dynamics.
The rf electric-field flipper may be applied in precursor experiments \cite{Lehrach:2012eg}. The initial beam polarization is planned to be vertical. The initial horizontal beam polarization can also be used.

The stimulating frequency, $\omega'$, should either (almost) coincide with that of the spin rotation, $\omega_0$, or differ by $n\omega_c~(n=\pm1,\pm2,\dots)$, where $\omega_c$ is the cyclotron frequency. This property can be properly substantiated and a rigorous quantitative description of the spin evolution can be given.

One usually puts devices like the electric-field flipper and the rf Wien filter into a straight section of the storage ring.
Since lengths of the flipper and the filter are small as compared with the ring circumference, an approximation of the perturbing field by the delta function is permissible. An expansion of the delta function into the Fourier series is defined by the known formula:
$$\sum^\infty_{n=-\infty}{\delta{(\Phi-2\pi n)}}=\frac{1}{2\pi}+\frac{1}{\pi}\sum^\infty_{n=1}{\cos{(n\Phi)}}=\frac{1}{2\pi}\sum^\infty_{n=-\infty}{\cos{(n\Phi)}},$$
where $\Phi=\omega_ct$ is the phase.

As a result, the following relations are valid (see, e.g., Ref. \cite{LeeFormula}):
\begin{equation}
\begin{array}{c}
\sin{(\omega' t+\chi)}\!\sum^\infty_{n=-\infty}{\delta{(\Phi-2\pi n)}}=\frac{1}{2\pi}\!\sum^\infty_{n=-\infty}{\sin{[(\omega'+n\omega_c)t+\chi]}}=\frac{1}{2\pi}\!\sum^\infty_{n=-\infty}{\sin{[(n+\nu)\Phi+\chi]}},\\
\cos{(\omega' t+\chi)}\!\sum^\infty_{n=-\infty}{\delta{(\Phi-2\pi n)}}=\frac{1}{2\pi}\!\sum^\infty_{n=-\infty}{\cos{[(\omega'+n\omega_c)t+\chi]}}=\frac{1}{2\pi}\!\sum^\infty_{n=-\infty}{\cos{[(n+\nu)\Phi+\chi]}},
\end{array}
\label{Lee}\end{equation}
where $\nu=\omega'/\omega_c$ is the modulation tune. In this case, $\omega_0=(n+\nu)\omega_c$ and $\omega'=(\nu_s+K)\omega_c$, where $\nu_s=\omega_0/\omega_c$ is the spin tune.

Equation (\ref{Lee}) shows a possibility to use resonance devices at different frequencies. In particular, an appropriate choice for the proton and the deuteron is $K=-2,-3$ and $K=+1,+2$, respectively.

More adequately, the electric fields of the flipper and the filter can be characterized as follows:
\begin{equation}
\begin{array}{c}
\bm \Omega_\|=2\bm{\mathfrak{E}}\cos{(\omega' t+\chi)},\qquad \bm{\mathfrak{E}}=-\frac{e\eta}{4m}\bm{\mathcal{E}}(\Phi),\\ 
\bm{\mathcal{E}}(\Phi)=\left\{\begin{array}{cc} \bm{E}_0 &\quad {\rm if} \quad \Phi\in \left[-\frac {\pi l}{C}+2\pi n,\, \frac {\pi l}{C}+2\pi n\right] \\ 0 &\quad {\rm if} \quad \Phi \notin \left[-\frac {\pi l}{C}+2\pi n,\, \frac{\pi l}{C}+2\pi n\right] \end{array}\right.\,,\qquad n=0,\pm1,\pm2,\dots,
\end{array}
\label{Leedefn}\end{equation} where $l$ is the length of the flipper/filter, $C$ is the ring circumference. The spin-dependent part of the classical Hamiltonian is defined by Eq. (\ref{propt}) and the electric field $\bm E_0$ is directed radially.

In this case, an expansion into the Fourier series has the form
\begin{equation}\bm{\mathcal{E}}(\Phi)=\bm{E}_0 
\sum^\infty_{n=-\infty}{a_n\cos{(n\Phi)}},\label{Leedenn}\end{equation}
where
\begin{equation}
\begin{array}{c} a_0=\frac{l}{C}, \qquad a_n=\frac{1}{\pi n}\sin{\frac{\pi n l}{C}}.
\end{array}
\label{LeedeFc}\end{equation} As a result,
\begin{equation}\begin{array}{c} 
\bm\Omega_\|=-\frac{e\eta}{2m}\bm{E}_0\sum^\infty_{n=-\infty}{a_n\cos{[(\omega'+n\omega_c)t+\chi]}}
=-\frac{e\eta}{2m}\bm{E}_0\sum^\infty_{n=-\infty}{a_n\cos{[(n+\nu)\Phi+\chi]}}.\end{array}\label{Leedemm}\end{equation}
Equations (\ref{LeedeFc}) and  (\ref{Leedemm}) show that the considered devices are not effective for oscillation modes $n>C/(2l)$. Otherwise, the Fourier coefficients for the delta function and the flipper/filter of a finite length agree on condition that $(n l/C)\ll1$. For a more precise Fourier expansion, one can use real parameters of the resonator fields.

The horizontal component of the angular velocity of the spin precession conditioned by a resonance interaction \emph{of the EDM} with the radial electric field can be presented in the form
\begin{equation}
\begin{array}{c}  \bm \Omega_\|=2\bm{\mathfrak{E}}\cos{(\omega t+\chi)}, \qquad \bm{\mathfrak{E}}=-\frac{e\eta}{4m}a_n\bm{E}_0.
\end{array}
\label{rfeq}\end{equation}
The resonance frequency, $\omega$, satisfies the condition \begin{equation}
\omega\equiv\omega'+n\omega_c\approx\omega_0.\label{rmode}\end{equation}

An expansion of a magnetic field in the rf Wien filter into the Fourier series is very similar. Evidently, a resonance influence of continuous and noncontinuous perturbing fields on the spin is practically the same.

The results presented
give an exhaustive description of storage ring resonance effects caused by the magnetic dipole
moment (MDM). For this purpose, one should simply substitute needed expressions for $\bm \Omega_\|$ and $\bm{\mathfrak{E}}$ into corresponding equations. Specifically,
\begin{equation}
\begin{array}{c} \bm \Omega_\|=2\bm{\mathfrak{E}}\cos{(\omega t+\chi)}=-\frac{e}{m}\left(G+\frac1\gamma\right)
a_n\bm{B}_0^{(r)}\cos{(\omega t+\chi)}
\end{array}
\label{Bradial}\end{equation}
for the radial magnetic field and
\begin{equation}
\begin{array}{c} \bm \Omega_\|=2\bm{\mathfrak{E}}\cos{(\omega t+\chi)}=-\frac{eg}{2m\gamma}
a_n\bm{B}_0^{(l)}\cos{(\omega t+\chi)}
\end{array}
\label{Blodial}\end{equation}
for the longitudinal magnetic field.
The resonance effect caused by the MDM and stimulated by the rf Wien filter with the radial magnetic and vertical electric fields [$(\bm E_0+\bm\beta\times\bm B_0^{(r)})_\|=0$] is defined by
\begin{equation}\begin{array}{c} \bm \Omega_\|=2\bm{\mathfrak{E}}\cos{(\omega t+\chi)}=-\frac{eg}{2m\gamma^2}
a_n\bm{B}_0^{(r)}\cos{(\omega t+\chi)}.
\end{array}
\label{Bfilter}\end{equation}

In Secs. \ref{ClaMres}--\ref{Quantum-mechanical}, the vertical direction of the stable spin axis is considered. The presence of the EDM tilts this direction. Therefore, the results obtained in Secs. \ref{ClaMres}--\ref{Quantum-mechanical} cannot be directly applied to describe the spin motion in storage ring EDM experiments. This problem is solved in the next section.

\section{Distinguishing features of a quasimagnetic resonance in storage ring electric-dipole-moment experiments}\label{Distinguishing}

Main distinguishing features of storage ring EDM experiments are a
simultaneous influence of external fields on the electric and
magnetic dipole moments and the existence of a resonance effect
even when the stimulating torque acting \emph{on the EDM} is equal
to zero. The last situation takes place when the resonance in a
EDM experiment is stimulated by the rf Wien filter with the
vertical magnetic and radial electric fields [$(\bm
E_0+\bm\beta\times\bm B_0^{(osc)})_r=0$]. The paradoxical property
of the existence of the resonance effect on condition that
$\bm{\Omega}_{EDM}=0$ has been first discovered by Semertzidis
\cite{SemertzidisRFE} with a computer simulation. The existence of
this effect has been confirmed and has been rigorously proven by
the subsequent theoretical analysis fulfilled in Refs.
\cite{OrlovPartiallyFrozenSpin,NikolaevRFE,rfWienFilter}.

In the present work, we give a very simple explanation of the distinguishing features of a quasimagnetic resonance in storage ring EDM experiments. This explanation is valid for any initial polarization of particles or nuclei.

Let us first consider a possibility of the resonance stimulated by the oscillating vertical magnetic field in the storage ring with the main magnetic field $\bm{B}_0$. The oscillating vertical magnetic field can be presented in the form $\bm{\mathcal{B}}(\Phi)\cos{(\omega 't+\chi)}$, where
\begin{equation}\bm{\mathcal{B}}(\Phi)=\bm{B}_0^{(osc)} 
\sum^\infty_{n=-\infty}{a_n\cos{(n\Phi)}}, \qquad \bm{B}_0^{(osc)}={B}_0^{(osc)}\bm e_z. \label{Leedelt}\end{equation}
An expansion into the Fourier series results in
\begin{equation}\bm{\mathcal{B}}(\Phi)\cos{(\omega 't+\chi)}=\bm{B}_0^{(osc)} 
\sum^\infty_{n=-\infty}{a_n\cos{[(\omega'+n\omega_c)t+\chi]}},\label{Leedetl}\end{equation}
where the coefficients $a_n$ are defined by Eq. (\ref{LeedeFc}). We consider only the resonance mode (\ref{rmode}).

The angular velocity of the spin rotation in the cylindrical coordinates is given by ($\bm\beta=\beta\bm e_\phi$)
\begin{equation} \bm\Omega^{(cyl)}=\omega_0\left[1+b_z\cos{(\omega t+\chi)}\right]\bm e_z- \frac{e\eta}{2m}\beta B_0\left[1+b_r\cos{(\omega t+\chi)}\right]\bm e_r, \qquad \omega_0=-\frac{eG}{m}B_0.
\label{eqtorll}\end{equation} We suppose that the quantities $\omega_0$ and $\beta$ can be positive and negative. In the considered case,
\begin{equation} b_r=b_r^{(m)}=b_z=b_z^{(m)}=\frac{a_nB_0^{(osc)}}{B_0}.
\label{eqtorff}\end{equation}

It is convenient to divide the vector
$\bm\Omega^{(cyl)}$ into the constant and oscillating parts, $\bm\Omega^{(0)}$ and $\bm\Omega^{(1)}$, as follows:
$$\bm\Omega^{(cyl)}=\bm\Omega^{(0)}+\bm\Omega^{(1)}\cos{(\omega t+\chi)}.$$
It is also convenient to consider separately the contributions from the electric and magnetic dipole moments. The former
contributions are proportional to $\eta$.

\begin{figure}[h]
\begin{center}
\includegraphics[width=4cm]{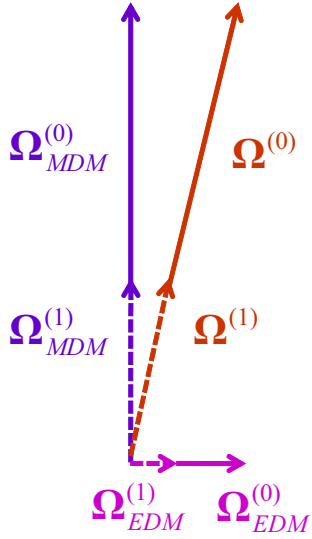}
\caption{Magnetic-field
``flipper''. A summation of the contributions from the EDM and the MDM leads to the collinearity of the constant and
oscillating parts of the vector $\bm\Omega^{(cyl)}$. A resonance effect does not exist.}
\label{fig1}
\end{center}
\end{figure}

Figure \ref{fig1} presents the angular velocity of the spin rotation.
Evidently, the constant and oscillating parts of the vector
$\bm\Omega^{(cyl)}$ are collinear. The vector $\bm\Omega^{(0)}$ defines the direction of the stable spin axis and forms the small angle
\begin{equation} \vartheta =\sin{\vartheta} =- \frac{e\eta}{2m\omega_0}\beta B_0=\frac{\eta\beta}{2G}
\label{eqtheta}\end{equation} with the $z$ axis. We may consider only terms linear in $\eta$. In this approximation,
\begin{equation} \bm\Omega^{(cyl)}=\omega_0\left[1+b_z\cos{(\omega t+\chi)}\right]\bm e_\vartheta, \qquad \bm e_\vartheta=\bm e_z+\vartheta \bm e_r.
\label{eqtorfn}\end{equation}
Equation (\ref{eqtorfn}) and Fig. \ref{fig1} show that any resonance effect does not exist.

It can be similarly proven that the resonance effect does not appear in an all-electric storage ring when main and oscillating electric fields are also radial.

The situation is different when the resonance is stimulated by the rf electric-field flipper in the storage ring with the main magnetic field. In this case,
\begin{equation} \bm E_0=E_0\bm e_r,\qquad b_r=b^{(e)}_r=\frac{a_nE_0}{\beta B_0},\qquad b_z=b^{(e)}_z=-\frac{\beta a_nE_0}{GB_0}\left(\frac{1}{\gamma^2-1}-G\right).
\label{eqtoref}\end{equation}
All parameters can be positive and negative.

It is very convenient to switch to the new axes, $\bm e_\zeta=\bm e_r-\vartheta \bm e_z,\,\bm e_\phi$ and $\bm e_\vartheta$, to describe the resonance spin dynamics in the general case. In this case,
\begin{equation} \bm\Omega^{(cyl)}=\omega_0\left[1+b_z\cos{(\omega t+\chi)}\right]\bm e_\vartheta+\omega_0\vartheta\left(b_r-b_z\right)\cos{(\omega t+\chi)}\bm e_\zeta.
\label{eqtornw}\end{equation}
As before, the vector $\bm e_\vartheta$ specifies the direction of the stable spin axis.
The parameters used are defined by Eqs. (\ref{eqtorll}), (\ref{eqtheta}), and (\ref{eqtoref}). It can be checked that
\begin{equation} b^{(e)}_r-b^{(e)}_z=\frac{a_nE_0}{\beta B_0}\cdot\frac{G+1}{G\gamma^2}=-\frac{ega_nE_0}{2m\beta\gamma^2\omega_0}.
\label{eqrmz}\end{equation}

\begin{figure}[h]
\begin{center}
\includegraphics[width=4cm]{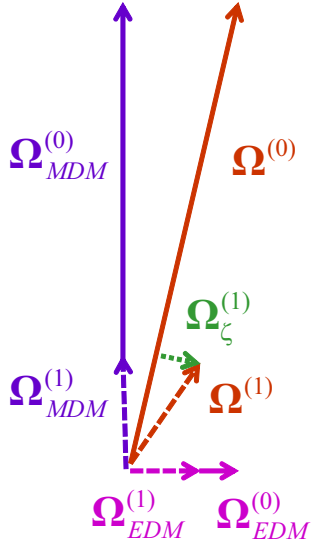}
\caption{Electric-field
flipper. The resonance effect takes place and is defined by the vector $\bm\Omega_\zeta^{(1)}$.}
\label{fig2}
\end{center}
\end{figure}

This situation is illustrated by Fig. \ref{fig2}. The vectors $\bm\Omega^{(0)}$ and $\bm\Omega^{(1)}$ are not collinear and the resonance effect takes place. 

If $|b_z|\ll1$, the general equations obtained in the precedent sections can be used. In this case,
\begin{equation} \mathfrak{E}=\frac12\omega_0\vartheta\left(b^{(e)}_r-b^{(e)}_z\right)
=-\frac{e\eta}{4m}\cdot\frac{G+1}{G\gamma^2}a_nE_0.
\label{pesonef}\end{equation}
When the initial spin direction is horizontal, the vertical spin polarization defined by Eq. (\ref{polvmol}) reads
\begin{equation}
\begin{array}{c}
P_z(t)=\mathfrak{E}t\sin{(\psi-\chi)}=-\frac{e\eta}{4m}\cdot\frac{G+1}{G\gamma^2}a_nE_0t\sin{(\psi-\chi)}.
\end{array}\label{polvmoe} \end{equation}

This equation agrees with the previous results \cite{OrlovPartiallyFrozenSpin,rfWienFilter}. We can mention that the longitudinal direction is opposite to $\bm e_\phi$ for positively charged particles. We underline that the present study provides for the description of the beam polarization in the general case. In particular, the horizontal spin polarization at the initial vertical spin direction is given by
\begin{equation}
\begin{array}{c}
P_x(t)=\mathfrak{E}t\sin{(\omega t+\chi)},\\
P_y(t)=-\mathfrak{E}t\cos{(\omega t+\chi)}.
\end{array}
\label{polvmov} \end{equation} Just the initial vertical beam
polarization will be used in the planned experiments at COSY
\cite{Lehrach:2012eg,RathmannWienFilter}.

As a rule, the condition $|b_z|\ll1$ is a satisfactory
approximation because the length of the flipper is much less than
the ring circumference ($l\ll C$). In the planned deuteron EDM
precursor experiment \cite{Lehrach:2012eg}, $l/C=5\times10^{-3}$.

The precedent studies
\cite{SemertzidisRFE,OrlovPartiallyFrozenSpin,NikolaevRFE,rfWienFilter}
have shown that the addition of the rf magnetic-field flipper with
the vertical field to the rf electric-field flipper significantly improves beam dynamics but does
not eliminate the EDM effect. This
conclusion first made thank to the computer simulation \cite{SemertzidisRFE} is confirmed by the geometrical method.

The above-mentioned addition allows one to obtain the rf Wien
filter and leads to a very unusual spin dynamics. The Lorentz force in this device vanishes. Evidently, the
resonance part of the radial component of $\bm\Omega^{(cyl)}$ is proportional to the Lorentz force ($\bm\beta\cdot\bm E=0$) and also vanishes. In this case, $B_0^{(osc)}=-E_0/\beta$. The angular
velocity of the spin rotation takes the form
\begin{equation} \bm\Omega^{(cyl)}=\omega_0\left[1+(b^{(e)}_z+b^{(m)}_z)\cos{(\omega t+\chi)}\right]\bm e_z
+\omega_0\vartheta\bm e_r, \label{eqtorwf}\end{equation} where
$\vartheta$ is given by Eq. (\ref{eqtheta}). Amazingly, there is
not an effect of the resonance field on the EDM and the resonance
effect \emph{proportional to the EDM} is ensured by the action of the
oscillating fields on the MDM. To determine this
effect, it is convenient to pass to the axes $\bm e_\vartheta$ and
$\bm e_\zeta$:
\begin{equation} \bm\Omega^{(cyl)}=\omega_0\left[1+\delta\cos{(\omega t+\chi)}\right]\bm e_\vartheta
-\omega_0\vartheta\delta\cos{(\omega t+\chi)}\bm
e_\zeta. \label{eqtorvf}\end{equation} It follows from Eqs. (\ref{eqtorll}), (\ref{eqtorff}) and the relation $b^{(m)}_r=-b^{(e)}_r$ that
\begin{equation} \delta=b^{(e)}_z+b^{(m)}_z=b^{(e)}_z-b^{(e)}_r=-\frac{a_nE_0}{\beta B_0}\cdot\frac{G+1}{G\gamma^2}.
\label{delta}\end{equation}

\begin{figure}[h]
\begin{center}
\includegraphics[width=4cm]{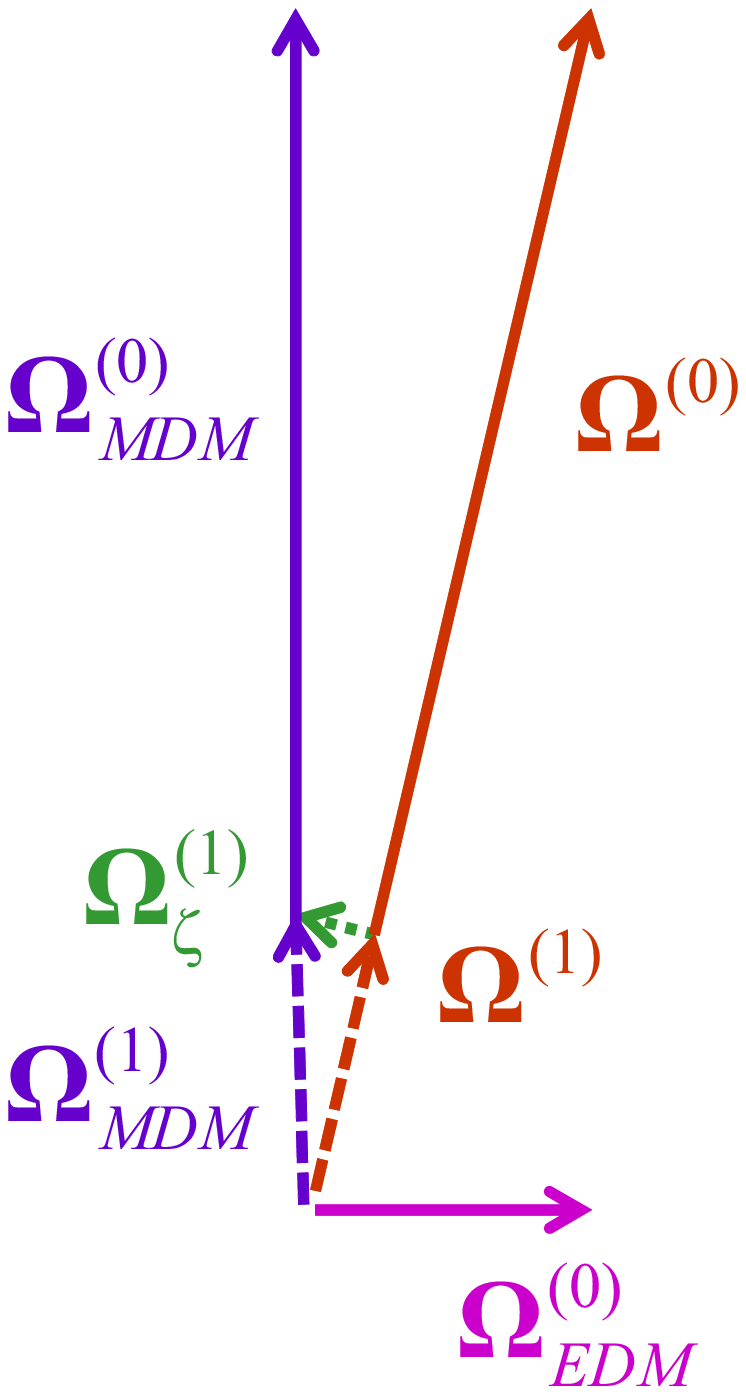}
\caption{rf Wien filter. While the resonance action on the EDM vanishes, the resonance effect
is ensured by the action of the
oscillating fields on the MDM and is defined by the vector $\bm\Omega_\zeta^{(1)}$.}
\label{fig3}
\end{center}
\end{figure}

Figure \ref{fig3} shows that the vectors $\bm\Omega^{(0)}$ and $\bm\Omega^{(1)}$ are not collinear and therefore the rf Wien filter provides for the resonance effect.

If we neglect $\delta$ as compared with unit in the term proportional to $\bm e_\vartheta$, we can
use the general equations obtained in the precedent sections. In
this case,
\begin{equation} \mathfrak{E}=-\frac12\omega_0\vartheta\delta
=-\frac{e\eta}{4m}\cdot\frac{G+1}{G\gamma^2}a_nE_0.
\label{pesonee}\end{equation}

In this approximation, an addition of the oscillating vertical
magnetic field does not change the resonance effect. However,
taking into account terms of the order of $\delta$ demonstrates a
difference between the EDM effects caused by the rf electric-field
flipper and the rf Wien filter (containing the same rf electric-field
flipper). While the difference is small, it
is crucial for the establishment of consent between analytical
derivations and computer simulations.

To simplify the derivation, we may consider the
case when $\chi=0,\,\omega=\omega_0$ and the initial beam polarization is azimuthal. In this case, the spin azimuth is equal to
$$\phi=\frac{\pi}{2}+\omega_0t+\delta\sin{\omega_0t}$$
and the azimuthal beam polarization is given by
\begin{equation} P_\phi(t)=\sin{\phi}=\cos{(\omega_0t+\delta\sin{\omega_0t})}.
\label{azimuth}\end{equation}

As a result, the change of the beam polarization along the $\bm e_\vartheta$ axis during one spin revolution reads
\begin{equation} \Delta P_\vartheta=-\omega_0\vartheta\delta\int_{-T/2}^{T/2}{\cos{(\omega_0t+\delta\sin{\omega_0t})}
\cos{\omega_0 t}\,dt}, \label{deriv}\end{equation} where
$T=2\pi/\omega_0$. Since
$$\begin{array}{c} \cos{(\omega_0t+\delta\sin{\omega_0t})}\cos{\omega_0 t}=\frac12\left[\cos{(\delta\sin{\omega_0t})}+
\cos{(2\omega_0t+\delta\sin{\omega_0t})}\right], \end{array}$$ we
can apply properties of the Bessel functions ($n$ is integer):
$$\begin{array}{c} J_n(z)=\frac{1}{\pi}\int_0^\pi{\cos{(nx-z\sin{x})}dx},\qquad J_{-n}(z)=(-1)^nJ_n(z). \end{array}$$
With the use of tables of integrals \cite{GradshteynRyzhik}, we
obtain the following \emph{exact} formula:
\begin{equation} \Delta P_\vartheta=\pi\vartheta[J_0(|\delta|)+J_2(|\delta|)]\delta.
\label{derivfl}\end{equation}

As a result, the average build-up of the vertical spin
polarization is given by
\begin{equation} \begin{array}{c}
P_z(t)=-\frac{\omega_0\vartheta\delta}{2}[J_0(|\delta|)+J_2(|\delta|)]
t=
-\frac{e\eta}{4m}\cdot\frac{G+1}{G\gamma^2}a_nE_0[J_0(|\delta|)+J_2(|\delta|)]
t.
\end{array}\label{polvmfn} \end{equation}
We can add that
$$J_0(|\delta|)+J_2(|\delta|)=\frac{2}{|\delta|}J_1(|\delta|).$$
When $|\delta|\ll1,~ J_0(|\delta|)+J_2(|\delta|)\approx1-\delta^2/8$.

When $\delta\rightarrow0$, we obtain the approximate equation
(\ref{polvmoe}) (on condition that $\psi=\pi/2+\chi$).

Similar equations 
are also valid for the rf
electric-field flipper. In this case,
\begin{equation} \begin{array}{c} \Delta P_\vartheta=\pi\vartheta[J_0(|b^{(e)}_z|)+J_2(|b^{(e)}_z|)]\delta,\\
P_z(t)=-\frac{\omega_0\vartheta\delta}{2}[J_0(|b^{(e)}_z|)+J_2(|b^{(e)}_z|)]
t=
-\frac{e\eta}{4m}\cdot\frac{G+1}{G\gamma^2}a_nE_0[J_0(|b^{(e)}_z|)+J_2(|b^{(e)}_z|)]
t.
\end{array}\label{derinll}\end{equation}

Thus, actions of the rf electric-field
flipper and the rf Wien filter on the spin can be distinguished while the difference is rather small. The exact equations (\ref{polvmfn}), (\ref{derinll}) should be applied for a comparison of
analytical results with computer simulations. Moreover, the use of the difference between the resonance effects conditioned by the rf electric-field flipper and the rf Wien filter is necessary for checking spin tracking programs.


One of key problems in EDM experiments is the problem of systematical errors. Equations (\ref{eq7}), (\ref{Nelsonh}) show that the vertical electric field and the radial and longitudinal magnetic fields may create a resonance effect imitating the presence of the EDM. This effect can take place due to misalignments and imperfections of the oscillating fields in the rf Wien filter. Similarly directed constant imperfection fields can also exist in the storage ring. However, they do not create any \emph{resonance} effect.

To decrease systematic errors, it is necessary to avoid any dependence of the particle motion on the fields of the rf Wien filter. This means canceling the Lorentz force in both radial and vertical directions \cite{rfWienFilter}. With allowance for Eq. (\ref{eq7}) and the relation $(\bm E_0+\bm\beta\times\bm B_0^{(osc)})\left|_z\right.=0$, we obtain the formula ($\bm\beta=\beta\bm e_\phi$)
\begin{equation}
\begin{array}{c}
\bm \Omega_\|=-\frac{e}{m}\cdot\frac{G+1}{\gamma^2}a_n\bm{B}^{(r)},\qquad
\mathfrak{E}=-\frac{e}{2m}\cdot\frac{G+1}{\gamma^2}a_nB_0^{(r)}.
\end{array}\label{polvmof} \end{equation}
This formula agrees with the result obtained in Ref. \cite{rfWienFilter}. A nonzero value of $B_0^{(r)}$ conditions a systematic error. It is rather difficult to distinguish the EDM signal from this systematic error. One can use the fact that these quantities differently depend on the velocity \cite{rfWienFilter}.

Another systematic error is caused by the longitudinal magnetic field. This systematic error has not been considered in Ref. \cite{rfWienFilter} but it has been mentioned later \cite{Rosenthal}. Equation (\ref{eq7}) shows that it causes the resonance effect defined by
\begin{equation}
\begin{array}{c}
\bm \Omega_\|=-\frac{e}{m}\cdot\frac{G+1}{\gamma}a_n\bm{B}^{(l)},\qquad
\mathfrak{E}=-\frac{e}{2m}\cdot\frac{G+1}{\gamma}a_n{B}_0^{(l)}.
\end{array}\label{polfmol} \end{equation}

One more systematical error can appear when the particle velocity inside the rf Wien filter has a vertical component. This component can be conditioned by field misalignments. In this case,
\begin{equation}
\begin{array}{c}
\bm \Omega_\|=\frac{e}{m}\left(G+\frac{1}{\gamma+1}\right)\beta_za_nE_0\cos{(\omega t+\chi)}\bm{e}_\phi,\qquad
\mathfrak{E}=\frac{e}{2m}\left(G+\frac{1}{\gamma+1}\right)\beta_za_nE_0.
\end{array}\label{polfmov} \end{equation}
This systematical error also consists in the spin rotation around the longitudinal direction.

In the presence of the longitudinal magnetic field or the vertical component of the particle velocity, the spin turns around the longitudinal direction while the EDM effect consists in the spin rotation
around the radial direction. As a result, phases of rotating horizontal spin components appearing due to the both 
above-mentioned sources of systematical errors and due to the EDM effect differ on $\pi/2$.

When the initial beam polarization is horizontal and the rf Wien filter is well-synchronized with the spin rotation, $\omega=\omega_0$ and the phase difference in Eqs. (\ref{findp}) and (\ref{polvmop}) is $\psi-\chi=\pm\pi/2$. In this case, the systematical errors caused by the longitudinal magnetic field and the vertical component of the particle velocity vanish and the only important systematical error is conditioned by the vertical electric field and the radial magnetic one.
However, the former systematical errors should be taken into account when the resonance conditions are not exactly satisfied.

\section{Summary}\label{summary}

In the present paper, a general theoretical description of the MR
is given. We have derived the general formulas describing a
behavior of all components of the polarization vector at the MR and have considered the case of an arbitrary initial polarization.
The equations obtained are exact on condition that the
nonresonance rotating field is neglected. The spin dynamics has
also been calculated at frequencies far from resonance without
neglecting the above-mentioned field. A quantum-mechanical
analysis of the spin evolution at the MR has been fulfilled and
the full agreement between the classical and quantum-mechanical
approaches has been proven.

Distinguishing features of magnetic
and quasimagnetic resonances for particles and nuclei moving in
accelerators and storage rings (including resonances caused by the
EDM) have been investigated in detail. We have considered the quasimagnetic resonance in a noncontinuous perturbing field. We have also fulfilled a detailed description of a quasimagnetic resonance in storage ring EDM experiments.
We have applied the simple geometrical method and have determined spin dynamics in the general case. We have shown for the first time
the difference between the resonance effects conditioned by the rf electric-field flipper and the rf Wien filter. The existence of this difference is crucial for the establishment of consent between analytical
derivations and computer simulations and for checking spin tracking programs.

The results obtained define also spin dynamics caused by systematical errors which appear due to misalignments and imperfections of the resonance fields in the rf Wien filter.

\section*{Acknowledgements}

The author is grateful to Y. Tsalkou for the help in graphic design and acknowledges the support by the Belarusian Republican Foundation for Fundamental Research
(Grant No. $\Phi$16D-004) and
by the Heisenberg-Landau program of the German Ministry for
Science and Technology (Bundesministerium f\"{u}r Bildung und
Forschung).


\begin{thebibliography}{}

\bibitem{Slichter}
C. P. Slichter, \emph{Principles of Magnetic Resonance,} 3rd edition (Springer-Verlag, Berlin, 1990)

\bibitem{MLevitt}
M. H. Levitt, \emph{Spin Dynamics: Basics of Nuclear Magnetic Resonance,} 2nd edition (Wiley, New York, 2008)

\bibitem{Thomas-BMT}
L. H. Thomas, The Motion of the Spinning Electron, Nature (London) \textbf{117}, 514 
(1926);
The Kinematics of an Electron with an Axis, Philos. Mag. {\bf 3}, 1 (1927); 
V. Bargmann, L. Michel, and V. L. Telegdi, Precession of the Polarization of Particles Moving in a Homogeneous Electromagnetic Field, Phys. Rev. Lett. {\bf 2}, 435 
(1959). This equation was also derived by J. Frenkel, Die Elektrodynamik des rotierenden Elektrons,
Z. Phys. {\bf 37}, 243 
(1926)

\bibitem{NKFS}
D. F. Nelson, A. A. Schupp, R. W. Pidd and H. R. Crane, Search for an Electric Dipole Moment of the Electron, Phys. Rev. Lett. {\bf 2}, 492 
(1959);
I. B. Khriplovich, Feasibility of search for nuclear electric dipole moments at ion storage rings,
Phys. Lett. B \textbf{444}, 98 
(1998)

\bibitem{GBMT}
T. Fukuyama and A. J. Silenko, Derivation of Generalized
Thomas-Bargmann-Michel-Telegdi Equation for a Particle with
Electric Dipole Moment, Int. J. Mod. Phys. A \textbf{28}, 1350147
(2013)

\bibitem{PhysScr}
A. J. Silenko, Spin precession of a particle with an electric
dipole moment: contributions from classical electrodynamics and
from the Thomas effect, Phys. Scripta \textbf{90}, 065303 (2015)

\bibitem{rfWienFilter}
W. M. Morse, Y. F. Orlov, and Y. K. Semertzidis, rf Wien filter in
an electric dipole moment storage ring: The partially frozen spin
effect, Phys. Rev. ST Accel. Beams \textbf{16}, 114001 (2013)

\bibitem{Mey}
S. Mey and R. Gebel, A Novel RF $\bm E \times\bm B$ Spin Manipulator at COSY, Int. J. Mod. Phys.: Conf. Ser. \textbf{40}, 1660094 (2016)

\bibitem{rfWienFilterNIMA}
J. Slim, R. Gebel, D. Heberling, F. Hinder, D. H\"{o}lscher, A. Lehrach, B. Lorentz, S. Mey, A. Nass, F. Rathmann, L. Reifferscheidt, H. Soltner, H. Straatmann, F. Trinkel, J. Wolters,
Electromagnetic Simulation and Design of a Novel Waveguide RF Wien
Filter for Electric Dipole Moment Measurements of Protons
and Deuterons, Nucl. Instrum. Meth. A \textbf{828}, 116 (2016) 

\bibitem{tracking}
E. M. Metodiev, I. M. D'Silva, M. Fandaros, M. Gaisser, S. Haci\"{o}mero\v{g}lu, D. Huang, K. L. Huanga,
A. Patil, R. Prodromou, O. A. Semertzidis, D. Sharma, A. N. Stamatakis, Y. F. Orlov, Y. K. Semertzidis,
Analytical benchmarks for precision particle tracking in electric and magnetic rings, Nucl. Instrum. Meth. A
\textbf{797}, 311 (2015) 

\bibitem{MuEDM08}
G. W. Bennett \emph{et al.} (Muon (\emph{g}$-$2) Collaboration), Improved Limit on the Muon Electric Dipole Moment, Phys. Rev. D \textbf{80},
052008 (2009)

\bibitem{PRDfinal}
G. W. Bennett \emph{et al.} (Muon (\emph{g}$-$2) Collaboration), Final report of the E821 muon anomalous magnetic moment measurement at BNL,
Phys. Rev. D \textbf{73}, 
072003 (2006)

\bibitem{Lehrach:2012eg}
  A.~Lehrach, B.~Lorentz, W.~Morse, N.~Nikolaev, and F.~Rathmann,
 Precursor Experiments to Search for Permanent Electric Dipole Moments (EDMs) of Protons and Deuterons at COSY,
  arXiv:1201.5773 [hep-ex]

\bibitem{RathmannWienFilter}
F. Rathmann, A. Saleev and N. N. Nikolaev, The search for electric dipole moments of light ions
in storage rings, J. Phys.: Conf. Ser. \textbf{447}, 012011 (2013)

\bibitem{Bar3}
V. G. Baryshevsky, Birefringence effect in the nuclear
pseudoelectric field of matter and an external electric field for
a deuteron (nucleus) rotating in a storage ring,
arXiv:hep-ph/0504064; About influence of the deuteron electric and
magnetic polarizabities on measurement of the deuteron EDM in a
storage ring, arXiv:hep-ph/0510158; Spin rotation of polarized
beams in high energy storage ring, arXiv:hep-ph/0603191; V. G.
Baryshevsky, A. A. Gurinovich, Spin rotation and birefringence
effect for a particle in a high energy storage ring and
measurement of the real part of the coherent elastic zero-angle
scattering amplitude, electric and magnetic polarizabilities,
arXiv:hep-ph/0506135

\bibitem{Bar4}
V. G. Baryshevsky, Rotation of particle spin in a storage ring with a polarized beam and measurement of the particle EDM, tensor polarizability and elastic zero-angle scattering amplitude, J. Phys. G: Nucl. Part. Phys. \textbf{35}, 035102 (2008)

\bibitem{PRC}
A. J. Silenko, Tensor electric polarizability of the deuteron in storage-ring experiments, Phys. Rev. C {\bf 75} (1), 014003 (2007)

\bibitem{PRC2008}
A. J. Silenko, Potential for measurement of the tensor magnetic polarizability of the deuteron in storage ring experiments, Phys. Rev. C {\bf 77},
021001(R) (2008)

\bibitem{PRC2009}
A. J. Silenko, Potential for measurement of the tensor polarizabilities of nuclei in storage rings by the frozen spin method,
Phys. Rev. C {\bf 80}, 
044315 (2009)

\bibitem{dEDMtensor} V. G. Baryshevsky and
A. J. Silenko, Potential for the measurement of the tensor electric and magnetic polarizabilities of the
deuteron in storage-ring experiments with polarized beams, J. Phys. Conf. Ser. \textbf{295} 
012034 (2011)

\bibitem{FW}
L.\,L. Foldy, S.\,A. Wouthuysen, On the Dirac Theory of Spin 1/2
Particles and Its Non-Relativistic Limit,
Phys. Rev. \textbf{78}, 29 (1950); A.\,J. Silenko, Foldy-Wouthyusen transformation and semiclassical limit for relativistic particles
in strong external fields, Phys. Rev. A \textbf{77}, 012116 (2008); General method of the relativistic Foldy-Wouthuysen transformation
and proof of validity of the Foldy-Wouthuysen Hamiltonian, Phys. Rev. A \textbf{91}, 022103 (2015)

\bibitem{CzJP}
A.J. Silenko, Investigation of spin dynamics of spin-1/2 and spin-1 particles near resonance with Hamilton approach,
Czech. J. Phys., Suppl. \textbf{56} , F245 
(2006)  

\bibitem{Nicolaevici}
N. Nicolaevici, Bouncing Dirac particles: compatibility between
MIT boundary conditions and Thomas precession, Eur. Phys. J. Plus \textbf{132}, 21 (2017)

\bibitem{RPJSTAB}
A. J. Silenko, Equation of spin motion in storage rings in the cylindrical coordinate system,
Phys. Rev. ST Accel. Beams \textbf{9}, 
034003 (2006)

\bibitem{JINRLettCylr}
A. J. Silenko, Comparison of spin dynamics in the cylindrical and Frenet-Serret coordinate systems,
Phys. Part. Nucl. Lett. \textbf{12}, 8 (2015)

\bibitem{PRD}
A. J. Silenko and O. V. Teryaev, Semiclassical limit for Dirac
particles interacting with a gravitational field. Phys. Rev. D
{\bf 71}, 064016 (2005)

\bibitem{PRD2007}
A. J. Silenko and O. V. Teryaev,  Equivalence principle and experimental tests of gravitational spin effects.
Phys. Rev. D {\bf 76}, 061101(R) (2007)

\bibitem{OFSgravity}
Y. Orlov, E. Flanagan, Y. Semertzidis, Spin rotation by Earth's gravitational field in a ``frozen-spin'' ring, Phys. Lett. A \textbf{376}, 
2822 (2012) 

\bibitem{OMS}
Y. F. Orlov, W. M. Morse, and Y. K. Semertzidis, Resonance Method of Electric-Dipole-Moment Measurements in Storage Rings, Phys. Rev. Lett. \textbf{96}, 214802 (2006)

\bibitem{JETPLet}
A. J. Silenko, Connection between beam polarization and systematical errors in storage ring electric-dipole-moment experiments,
JETP Lett. \textbf{98}, 191 (2013)

\bibitem{FJM}
F. J. M. Farley, K. Jungmann, J. P. Miller, W. M. Morse, Y. F.
Orlov, B. L. Roberts, Y. K. Semertzidis, A. Silenko, and E. J.
Stephenson, A new method of measuring electric dipole moments in
storage rings, Phys. Rev. Lett. \textbf{93}, 052001 (2004)

\bibitem{SemertzidisRFE}
Y. K. Semertzidis, RFE and RFB effects, Storage Ring EDM Note
with simulation tracking (2012), http://www.bnl.gov/edm/files/pdf/YkS\_two\_RF\_2012\_0208.pdf

\bibitem{NikolaevRFE}
N. N. Nikolaev, Duality of the MDM-transparent
RF-E flipper to the transparent RF Wien-filter at all
magnetic storage rings, Storage Ring EDM Note (2012), http://www.bnl.gov/edm/files/pdf/NNikolaev\_Wien\_RFE.pdf

\bibitem{OrlovPartiallyFrozenSpin}
Y. F. Orlov, On the partially-frozen-spin method, Storage Ring EDM
Note (2012), http://www.bnl.gov/edm/files/pdf/YOrlov\_On\_partially-frozen-spin\_3\_21\_12.pdf

\bibitem{LeeFormula} S. Y. Lee, Spin resonance strength of a localized rf magnetic field,
Phys. Rev. ST Accel. Beams \textbf{9}, 074001 (2006)

\bibitem{GradshteynRyzhik}
I. S. Gradshteyn and I. M. Ryzhik, \emph{Table of Integrals,
Series, and Products.} 7nt edition, (Academic Press, Amsterdam,
2007)

\bibitem{Rosenthal}
M. Rosenthal, Investigation of Beam and Spin Dynamics for EDM Measurements at
COSY, Microscopy and Microanalysis \textbf{21}, Iss. S4, 30 (2015)   

\end{thebibliography}
\end{document}